\documentstyle[12pt]{article}

\begin{document}
\title{\bf{Trace anomaly and Casimir effect}}
\author{
M.R.Setare\footnote{E-mail:Mreza@physics.sharif.ac.ir} and
A.H.Rezaeian \footnote{E-mail:Rezaeian@physics.sharif.ac.ir}\\
\small{Department of Physics,Sharif University of Technology
Tehran-Iran}}
\date{\small{\today}}
 \maketitle

\begin{abstract}
The Casimir energy for scalar field of two parallel conductors in
two dimensional domain wall background, with  Dirichlet boundary
conditions,
 is calculated by making use of general properties of renormalized
 stress tensor. We show that vacuum expectation values of stress tensor contain
 two terms which come from the boundary conditions and the gravitational
 background. In two dimensions the minimal coupling reduces to the
 conformal coupling and stress tensor can be obtained by the local
 and non-local contribution of the anomalous trace. This work shows
 that there exists a subtle and deep connection between Casimir effect and
 trace anomaly in curved space time.
 \end{abstract}
\newpage
 \section{Introduction}
 In the semiclassical approximation theory of quantum gravity we are involved
 with calculation the expectation value of energy momentum tensor in special
 vacuum,  \cite{quantum}. However the usual expression
 for the stress tensor includes singular products of the field
 operators for stress tensor.Renormalization theory of the stress
 tensor claims to solve this problem, but it must be mentioned that
 the usual scheme of renormalization includes complexity and somewhat
 ambiguity. For instance, there is no conceptual support for a local
 measure of energy momentum of some given state without any reference
 to any global structure. We know in this frame energy is source of
 gravity and we are not allowed to subtract any unwanted part of
 energy even though it is infinite. So to consider the back reaction effect of the
 quantum field on the gravitational field,  we must
 find a more elaborate renormalization scheme in which the dynamics of
 gravitational field is a vital component.
 In original Casimir effect discovered in 1948 by H.B.G Casimir \cite{casimir}
 we are concerned about force and energy, but we are not usually interested in
 dynamics of the gravitational field. Even in many cases in
 curved boundary  problems, energy is not our main concern.
  Because of unphysical nature of boundary condition
  the energy diverges approaching to curved boundary \cite{deu-can}.
  The Casimir effect can be viewed as a polarization of vacuum by boundary conditions and
  external fields, such as gravitational field. In the present paper we are going
  to consider a simple example in which these two types of sources for vacuum
  polarization are present.There is several
  methods for calculating Casimir energy. For instance,  we can mention mode summation,
  Green's function method \cite{Plunien}, heat kernel method \cite{bor}along with appropriate
   regularization schemes such as point separation \cite{chr},\cite{adler}
  dimensional regularization \cite{deser}, zeta function regularization
  \cite{haw}. But it must be remarked that practically all of the
  methods are successful only for boundary conditions with high
  symmetry in flat space time. In fact we don't have any general procedure for
  renormalizing stress tensor in gravitational background with arbitrary boundary.

  In this paper the Casimir stress tensor for scalar field in two dimensional
  analog of domain wall space time for two parallel conductor plates
   with Dirichlet boundary conditions, is calculated.
  The Casimir stress tensor is obtained by imposing only general
  requirements which is discussed in section 2.We show direct
  relation between trace anomaly and Casimir effect,although we have been aware of
  role of anomalous trace in gravitational background such as
  Hawking effect \cite{chris}.

  \section{General properties of stress tensor}
  In semiclassical framework for yielding a sensible theory of back
  reaction Wald \cite{wald} has developed an axiomatic approach.
  There one tries to obtain an expression for the renormalized
  $T_{\mu\nu}$ from the properties (axioms) which it must fulfill.
  The axioms for the renormalized energy momentum tensor are as follow:

  1-For off-diagonal elements standard results should be obtained.

  2-In Minkowski space time standard results should be obtained.

  3-Expectation values of energy momentum are conserved.

  4-Causality holds .

  5-Energy momentum tensor contains no local curvature tensor depending on
  derivatives of the metric higher than second order.

  Two prescriptions that satisfy the first four axioms can differ
  at most by a conserved local curvature term.Wald, \cite{wald2}, showed any prescription
  for renormalized $T_{\mu \nu}$ which is consistent with axioms 1-4 must yield the
  given trace up to the addition of the trace of conserved local curvature .It
  must be noted  (that trace anomalies in stress-tensor,that is,the nonvanishing
  $T^\mu _\mu$ for a conformally invariant field after
  renormalization)
  are originated from some quantum behavior \cite{jac}. In two dimensional
  space time one can show that a trace-free stress tensor can not be
  consistent with conservation and causality
  if particle creation occurs.A trace-free ,conserved stress
  tensor in two dimensions must always remain zero if it is
  initially zero.
  One can show that the "Davies-Fulling-Unruh" \cite{Davies} formula
  for the stress tensor of scalar field which yield an anomalous trace
  ,$T^{\mu} _\mu=\frac{R}{24\pi}$, is the unique one which is
  consistent with the above axioms.
  In four dimensions, just as in two dimensions, a trace-free stress tensor which
  agrees with the formal expression for the matrix elements between orthogonal
  states can not be compatible with both conservation laws and causality .
  It must be noted that, as showed Wald\cite{wald2}, with Hadamard regularization
  in massless case axiom(5) can not be satisfied unless we
  introduce a new fundamental length scale for nature. Regarding all these
   axioms,thus, we are able to get an unambiguous
  prescription for calculation of stress tensor.
  \section{Vacuum expectation values of  stress tensor}
  It has been shown in \cite{Vil1},\cite{Vil2} that the
  gravitational field of the vacuum domain wall with a source of
  the form
  \begin{equation}
  T^{\nu}_{\mu}(x)=\sigma \delta(x)diag(1,0,1,1)
  \end{equation}
  does not correspond to any exact static solution of the
  Einstein's
  equation (for domain wall solution of  Einstein-scalar equation
  see\cite{Widrow}). However a static solution of the Einstein's
  equation representing a planar domain wall in an anisotropic
 background has been found \cite{man}. This solution matches in the
 weak-field region to the linearized solution of Vilenkin. The
 energy momentum tensor of the background has the form
 \begin{equation}
 T^{\mu}_{\nu}=\frac{\alpha(\alpha-2\gamma) k^{2}(1+kx)^{2(\alpha+\gamma)}}{8\pi
 G}diag(1,\frac{-3\alpha}{2\gamma-\alpha},1,1)
 \end{equation}
 For the energy density of the background to be positive we must
 have $\gamma<\frac{\alpha}{2}$, then we have the following Ricci
 tensor
 \begin{equation}
 R^{\mu}_{\nu}=\alpha(\gamma-2\alpha)k^{2}(1+kx)^{2(\alpha+\gamma)}diag(1,\frac{3\gamma}{\gamma-2\alpha},1,1)
 \end{equation}
 In this case we have the metric
    \begin{equation}
  ds^{2}=(1+kx)^{-2\alpha} (dt^{2}-dy^{2}-dz^{2})-(1+kx)^{-2(\alpha+\gamma+1)} dx^{2}
  \end{equation}
 where $\alpha ,\gamma, k>0$, This corresponds to a domain
 wall with surface tension $\sigma=\frac{\alpha k}{2\pi G}>0$. The
 space time on the domain wall is flat, and energy momentum tensor
 of the wall is invariant with respect to Lorentz boost in
 the $y-z$ plane.The above metric representing a gravitational
 field which is homogeneous and isotropic in the $y-z$ plane and
 has reflection symmetry with respect to the wall. In fact(4) is
 the special form of the following metric
 \begin{equation}
 ds^{2}=(1+k|x|)^{-2\alpha}dt^{2}-(1+k|x|)^{-2(\alpha+\gamma+1)}dx^{2}+
 (1+k|x|)^{-2\beta}(dy^{2}+dz^{2})
 \end{equation}
 The case $\beta=0$ and $\gamma=\alpha$ corresponds to the flat
 Kanser metric in Taub coordinates \cite{Amu}. In metric (4) we
 have $\alpha=\beta$.
   Now, just for the sake of simplicity, we consider two
   dimensions in which
  \begin{equation}
  d^{2}s=(1+kx)^{-2\alpha}(d^{2}t-d^{2}X)
  \end{equation}
  where
  \begin{equation}
  X=\frac{1-(1+kx)^{-\gamma}}{k\gamma}
  \end{equation}
  Let's define
  \begin{equation}
  \Omega(x)=(1+kx)^{-2\alpha}
  \end{equation}
  and so we have
  \begin{equation}
  \frac{dx}{dX}=\Omega(x)^{k'}        \hspace{2cm}
  k'=\frac{-(\gamma+1)}{2\alpha}
  \end{equation}
  From now on, our main goal is to determine for a general form
  of conserved energy momentum tensor, regarding trace anomaly for
  the metric $(4)$. For the non -zero Christoffel symbols of the
  metric $(4)$we have in (t,X) coordinate;
  \begin {equation}
  \Gamma^{X}_{tt}=\Gamma^{t}_{tX}=\Gamma ^{t}_{XX}=\Gamma^{X}_{XX}=\frac{\Omega^{k'-1}(x)}{2}
  \frac{d\Omega(x)}{dx}
  \end{equation}
   Then the conservation equation takes the following form
  \begin{equation}
  \partial_X{T^{X}_{t}}+\Gamma^{t}_{tX}T^{X}_{t}-\Gamma^{X}_{tt}T^{t}_{X}=0
  \end{equation}
  \begin{equation}
  \partial_{X}T^{X}_{X}+\Gamma^{t}_{tX}T^{X}_{X}-\Gamma ^{t}_{tX}T^{t}_{t}=0
  \end{equation}
  in which,
  \begin{equation}
  T^{t}_{X}=-T^{X}_{t}         \hspace{2cm}
  T^{t}_{t}=T^{\beta}_{\beta}-T^{X}_{X}
  \end{equation}
  and $T^{\beta}_{\beta}$ is anomalous trace in two dimension.Using the
  equations $(10-12)$ it could be shown that
  \begin{equation}
  \frac{d( T^{X}_{t}\Omega(x)^{k'})}{dx}=0
  \end{equation}
  and
  \begin{equation}
  \frac{d(\Omega^{k'}(x)T^{X}_{X})}{dx}=(1/2 \Omega^{k'-1}(x))
  (\frac{d\Omega(x)}{dx})T^{\beta}_{\beta}
  \end{equation}
  Then equation $(14)$leads to:
  \begin{equation}
  T^{X}_{t}=\alpha' \Omega^{-k'}(x)
  \end{equation}
  where $\alpha'$ is the constant of integration.The solution of Eq.(15) might be
  written in the following form
  \begin{equation}
  T^{X}_{X}(x)=(H(x)+\eta)\Omega^{-k'}(x)
  \end{equation}
  where
  \begin{equation}
  H(x)=1/2 \int^{r}_{l}T^{\beta}_{\beta}(x')\Omega^{k'-1}(x')\frac{d\Omega(x')}{dx'}dx'
  \end{equation}
  With $l$ being an arbitrary scale of length and considering
  \begin{equation}
  T^{\beta}_{\beta}=\frac{R}{24\pi}=\frac{\alpha(\gamma-2\alpha)k^2(1+kx)^{2(\alpha+\gamma)}}{12\pi}
  \end{equation}
  the function $H(x)$ produces the non-local contribution of the trace
  $T^{\beta}_{\beta}(x)$ to the energy momentum tensor. Finding
  $l$ depends on the metric. For the metric $(4)$ we choose $l=0$(to
  be definite we shall consider right half space of domain wall
  geometry) because we want to incorporate the non-local effect of trace
  anomaly in all space time,  so we reach
  \begin{equation}
  H(x) =\frac{(2\alpha-\gamma)\alpha^{2}k^{2}((kx+1)^{2\alpha+3\gamma+1}-1)}{12\pi(2\alpha+3\gamma+1)}
  \end{equation}
  Using the equations $(13),(16)$and $(17)$ it can be shown that
   energy momentum tensor takes the below form in (t,X)coordinates. So
   we have most general form of stress tensor field in our
   interesting background.
   \begin{equation}
   T^\mu\   _\nu(x)=\left(\begin{array}{cc}
   T^{\beta}_{\beta}-\Omega(x)^{-k'}H(x) & 0 \\
   0 &\Omega(x)^{-k'}H(x) \
   \end{array}\right)+\Omega^{k'}\left(\begin{array}{cc}
   -\eta&-\alpha'  \\
   \alpha' & \eta \
   \end{array}\right)
   \end{equation}

     Now we are going to obtain two constants  $\alpha^{'}$
     and $\eta$ by imposing the second axiom of
   renormalization scheme. So when we put $k=0$,  we reach the
   special case of Minkowski space time. The type of boundary
    condition which we choose is Dirichlet
    $\phi(x_{1})=\phi(x_{2})=0$.
     The standard Casimir stress in Minkowski
    space time is as follows
  \begin{equation}
  reg<0|T_{\mu \nu}|0>=\frac{-\pi}{24a^2}\delta_{\mu \nu}
  \end{equation}
  where $a$ is proper distance between the plates. For example, it could be
  obtained by mode summation using Abel-Plana formula \cite{aram}, without inserting
  any cut off. Comparing to $(21)$we obtain
  \begin{equation}
  \eta=\frac{\pi}{24a^2}      \hspace{2cm}
  \alpha^{'}=0
  \end{equation}
 Thus we have obtained the energy momentum tensor as direct sum of two
  terms;boundary term(second term) and term which presents the
  vacuum polarization in gravitational background in
  absence of boundaries(first term).
  \begin{equation}
  <T^\mu  _\nu>=<T^{(g) \mu}  _\nu>+<T^{(b)\mu}  _\nu>
  \end{equation}

  where $<T^{(g) \mu}  _\nu>$ and $<T^{(b)\mu}  _\nu>$ stand for gravitational
 and boundary parts respectively. It should be noted that trace anomaly
  has contribution in first term which comes from background not
  boundary effect.  However it has contribution in total Casimir
  energy momentum tensor.
 As in four dimension previously has been shown\cite{Setare} in the regions $x<x_{1}$and $x>x_{2}$the boundary part is zero
  and only gravitational polarization part is present,it is clear
  that the forces acting on plates are determined only by boundary
  part,when the effective pressure created by gravitational part
  is zero.
  \section{Conclusion}
  We have found the renormalized energy momentum tensor for scalar field
  with Dirichlet boundary conditions in domain wall
  background,  only by making use of general properties of stress tensor.
  It is in close relation to what is done
  in\cite{chris},\cite{ghafa}.  We propose that if we know the stress tensor
  for a given boundary
  in Minkowski space time, Casimir effect in gravitational background
  can be calculated.The result contains two parts, one comes from
  boundary conditions and the other one comes from the effect of
  gravitational background over the vacuum of scalar field,this part carries
  the local and nonlocal contributions of anomalous trace in
  complete Casimir effect in curved background.

\vspace{3mm}

{\large {\bf  Acknowledgement }}
 \vspace{1mm}
\small

Our special gratitude goes to Dr.H.Salehi.In addition we would
like to thank Mr.A.T.Rezakhani for helping us to prepare this
Latex format of the paper.

\end{document}